\documentclass{egpubl}

 \ConferenceSubmission 
 \electronicVersion 

\ifpdf \usepackage[pdftex]{graphicx} \pdfcompresslevel=9
\else \usepackage[dvips]{graphicx} \fi

\PrintedOrElectronic

\usepackage{t1enc,dfadobe}
\usepackage{mathptmx}
\usepackage{times}
\usepackage{subfigure}
\usepackage{amsmath}
\usepackage{amssymb}
\usepackage{mathtools}
\usepackage{algorithmic} 
\usepackage{algorithm}
\usepackage{amsfonts}
\usepackage{egweblnk}

\title%
      {Glyph Sorting: Interactive Visualization for Multi-dimensional Data}

\author[D.\ H.\ S. Chung \textit{et al.}]
       {David H. S. Chung$^{1,2}$, Philip A. Legg$^{1,2}$, Matthew L. Parry$^{1,2}$, Rhodri Bown$^{3}$, Iwan W. Griffiths$^{2}$, Robert S. Laramee$^{1}$ and Min Chen$^{4}$
        \\
         $^1$Department of Computer Science, Swansea University, UK.\\
         $^2$College of Engineering, Swansea University, UK.\\
         $^3$Centre of Excellence, Welsh Rugby Union, UK.\\
         $^4$Oxford e-Research Centre, Oxford University, UK.
       }

\begin{document}

\maketitle

\begin{abstract}

Glyph-based visualization is an effective tool for depicting multivariate information. 
Since sorting is one of the most common analytical tasks performed on individual attributes of a multi-dimensional data set, this motivates the hypothesis that introducing glyph sorting would significantly enhance the usability of glyph-based visualization.
In this paper, we present a glyph-based conceptual framework as part of a visualization process for interactive sorting of multivariate data.
We examine several technical aspects of glyph sorting and provide design principles for developing effective, visually sortable glyphs.
Glyphs that are visually sortable provide two key benefits: 1) performing comparative analysis of multiple attributes between glyphs and 2) to support multi-dimensional visual search.
We describe a system that incorporates focus and context glyphs to control sorting in a visually intuitive manner and for viewing sorted results in an Interactive, Multi-dimensional Glyph (IMG) plot that enables users to perform high-dimensional sorting, analyse and examine data trends in detail. 
To demonstrate the usability of glyph sorting, we present a case study in rugby event analysis for comparing and analysing trends within matches.
This work is undertaken in conjunction with a national rugby team. 
From using glyph sorting, analysts have reported the discovery of new insight beyond traditional match analysis.

\end{abstract}

\section{Introduction}
\label{sec:introduction}

Sorting large, multi-dimensional data is a growing consensus in modern data acquisition and processes where the ordering of data is an integral part of many applications and disciplines, ranging from the analysis of scientific information (e.g., using graphs and charts), to enhancing the efficiency of algorithms. 
Such records are traditionally sorted analytically in a data-driven manner (e.g., via spreadsheets), where users perform sorting on individual attributes of a multi-dimensional data set.
This is a non-trivial task due to the vast possible permutations of sorting which greatly impacts the expressiveness in high dimensional visualizations~\cite{yang03}.
When data must be ordered using a high level of sorting, it reveals two important challenges: 1) how the data is organised, and 2) the ordering of sort keys, which can not be easily observed by viewing large tables of data.

\begin{figure*}
	\centering
	\subfigure[]{
		\includegraphics[width = 3.5cm]{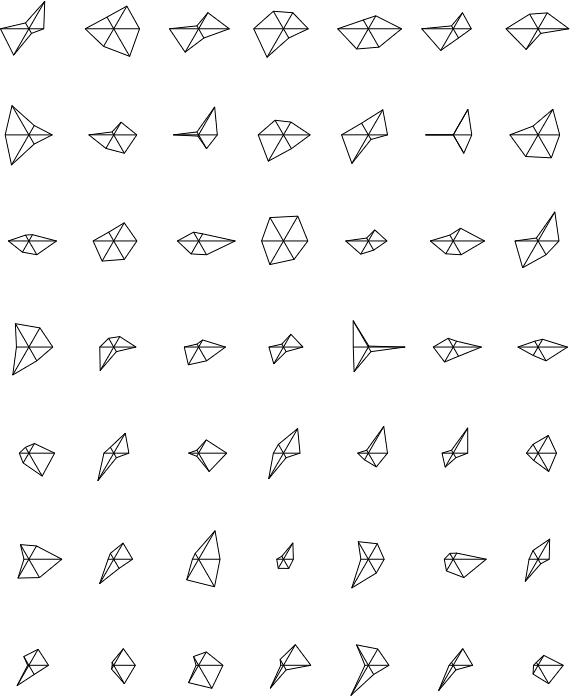} \hspace{0.5cm}
		\includegraphics[width = 3.5cm]{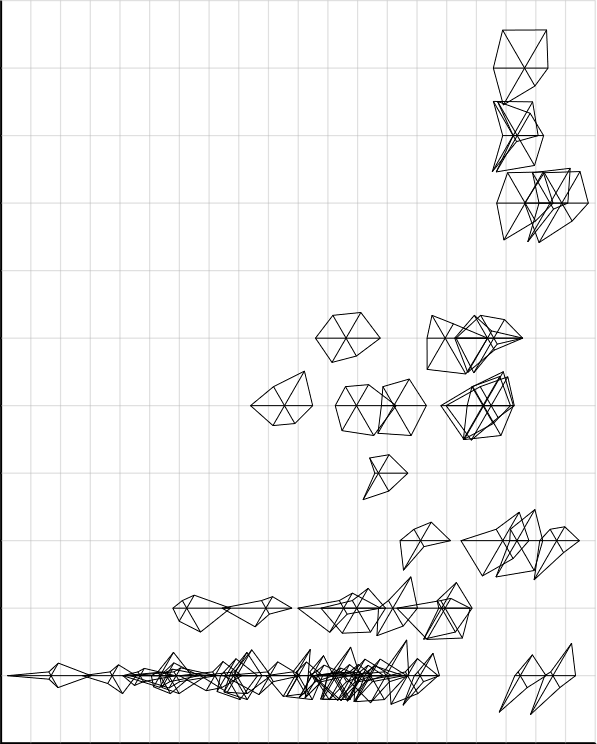}
		\label{fig:starGlyph}
	}
	\hspace{0.2cm}
	\subfigure[]{
		\includegraphics[width = 3.5cm]{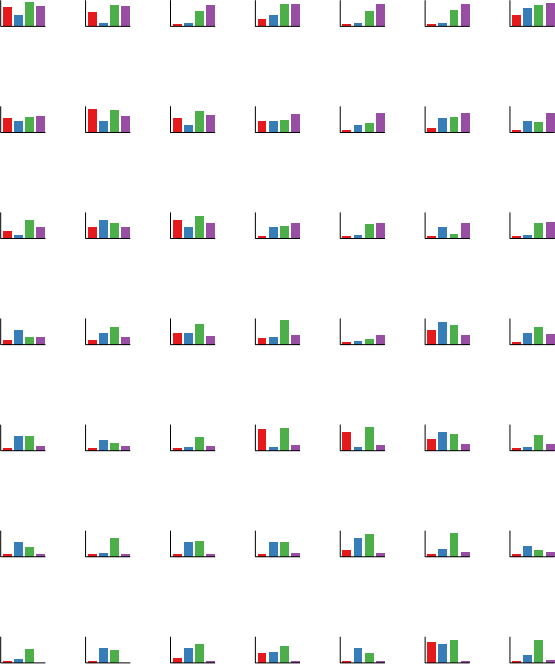} \hspace{0.5cm}
		\includegraphics[width = 3.5cm]{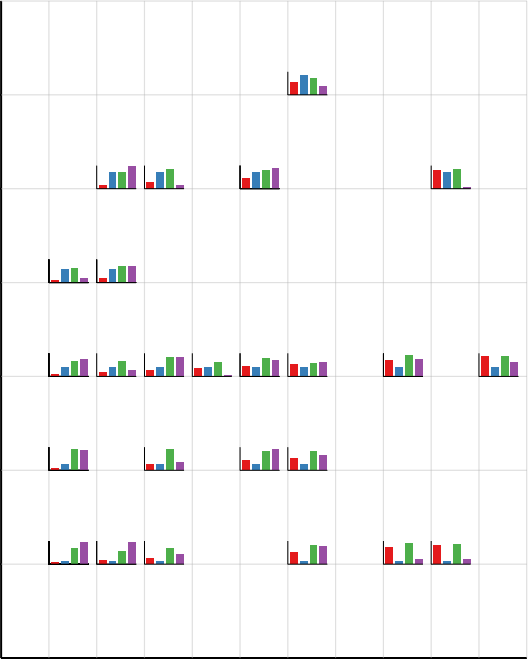}
		\label{fig:barChartGlyph}
	}
	\vspace*{-3mm}
	\caption{Visual representation of two example multi-dimensional glyphs, namely (a) Star glyphs and (b) Bar chart glyphs when glyphs on the left are unordered, in comparison to glyphs on the right which are ordered to two sorting parameters.}	 
\label{fig:visualSorting}
\vspace*{-6mm}
\end{figure*}

Glyphs (sometimes known as icons) are graphical entities that convey one or more data values using visual features such as size, shape and colour.
This significantly improves perception of data characteristics and is well suited for depicting high-dimensional, multivariate data~\cite{ward02}.
Chernoff Faces~\cite{chernoff73} and Star Glyphs~\cite{siegel72} are some examples of multivariate glyphs where identifying glyphs with similar features is effective, but cognitively challenging when determining the ordering of glyphs.
Thus, such glyphs are not visually sortable in an obvious way. 
This becomes a greater challenge when glyphs are unorganised. 
Figure~\ref{fig:visualSorting} demonstrates how ordering such glyphs in a given spatial configuration is more informative in revealing multivariate trends.
Glyph sorting is one approach for performing interactive sorting of multivariate data as part of a visualization process.
As an data exploration mechanism, interactive sorting in visualization provides the following additional objectives: 1) making observations about data patterns (e.g., clusters and distributions) in relation to a sorted variable and stimulating hypotheses about other variables. 2) performing analytical tasks and visual evaluation of hypotheses, such as what variables may affect the ordering of a specific variable.

In this paper, we present a novel glyph-based sorting framework to drive and facilitate interactive sorting of data in a visual and intuitive manner.
We describe a set of design principles (Section~\ref{sec:designprinciples}) for mapping attributes to visually sortable glyphs. 
This significantly enhances the usability of glyph-based visualizations for both comparative analysis of multi-variate data and for supporting visual search. 
In Section~\ref{sec:interactivesystem}, we present an interactive system for the exploration of glyph-based visualization. 
Novel features of the system include a focus and context glyph-based user interface (Section~\ref{sec:interface}) to control high-dimensional sorting and viewing sorted results in a \textit{Interactive, Multi-dimensional Glyph} (IMG) plot (Section~\ref{sec:IMGplot}).
We extend traditional axis mapping using hierarchical axis binning (Section~\ref{sec:axisbinning}). This enables visual depiction of multiple sort key parameters in space, which is effective for reducing visual clutter in the IMG plot view.
To demonstrate the effectiveness of glyph-sorting, we present a real-world case study of rugby event analysis.
The work is carried out in close collaboration with an international rugby team, in which we developed a glyph-sorting software tool for use by the coaching analysts. As a result of glyph sorting, the analysts uncovers new insight and knowledge for match analysis.
The main contributions of this paper are:

\begin{itemize}
\item The introduction and development of high-dimensional, focus and context glyphs that are visually sortable to support sorting of multi-variate data.
\item A novel glyph-based, interactive system for controlling high-dimensional sorting and viewing sorted results.
\item A hierarchical axis binning method for encoding multiple dimensions onto a single axis. This effectively reduces visual clutter by relaxing the positioning of glyphs.
\item An evaluation of the effectiveness of glyph sorting in a real-world case study of sports event analysis.  
\end{itemize}


\section{Related Work}
\label{sec:relatedwork}

Sorting is the computational process of rearranging a sequence of items into ascending or descending order~\cite{knuth73}. 
Many sorting algorithms have been proposed, including bubble sort by Demuth~\cite{demuth56}, merge sort by von Neumannr~\cite{knuth73}, and quick sort by Hoare~\cite{hoare62}.
Since best and worse case performance runtime can vary drastically with such algorithms, further research continues to propose new sorting techniques~\cite{bender06} and adaptive approaches that utilise ordered data~\cite{estivill-castro92}. 
Our work is not focused on a faster sorting algorithm per say, but combining the benefits of sorting with glyph-based visualization.

Glyph-based visualization is an established technique for depicting multi-dimensional data sets.
The survey by Ward~\cite{ward02, ward08} provides a technical framework for glyph-based visualization, covering aspects of visual mapping and layout methods, as well as addressing important issues such as bias in mapping and interpretation. 
Ropinski~\textit{et al.}~\cite{ropinski08} present an in-depth survey on the use of glyph-based visualization for spatial multivariate medical data. Glyphs are widely used in other application areas, such as DT-MRI visualization~\cite{laidlaw98mousespinal, westin02processing}, unsteady flow visualization~\cite{hlawatsch11flowRadar} and activity recognition~\cite{botchen08ActionBasedVideo}. Lie~\textit{et al.}~\cite{lie09criticaldesign} describe a general pipeline for visualizing scientific data in 3D using glyphs and introduce design guidelines such as the orthogonality of individual attribute mappings.
Pearlman~\textit{et al.}~\cite{pearlman07} use a glyph-based multivariate visualization to understand depth and diversity of large data sets.
Chlan and Rheingans~\cite{chlan05} use 2D and 3D glyph-based multivariate visualization to show distribution within the data set.
J\"{a}nicke~\textit{et al.}~\cite{heike10soundRiver} introduce \textit{SoundRiver}, that depict audio/video events from movies using glyphs for visualization on a timeline.
Previous to this study, Legg~\textit{et al.}~\cite{legg12} conducted a design study to show the effective use of glyph-based visualization within sports performance analysis.
A fundamental difference here is that we use glyphs that are visually sortable.

Interactive visualization studies the ability of human interaction for exploring and understanding datasets through visualization, which Zudilova~\text{et al.}~\cite{zudilova08} covers in a state-of-the-art report. 
De Leeuw and Van Wijk~\cite{deLeeuwVanWijk93probe} is one earlier research which incorporates glyphs into interactive visualization for analysing multiple flow characteristics in selected regions using a probe glyph.
Shaw~\textit{et al.}~\cite{shaw98} describe an interactive glyph-based framework for visualizing multi-dimensional data, where attributes are mapped in order of data importance to visual cues such as location, size, colour and shape. 
To our knowledge, this is the first work of its kind to introduce focus and context glyphs for visual sorting of high-dimensional data.


\section{Sorting: Entities and Sort Keys}
\label{sec:sortingdefinition}

Sorting is the most common analytical task which is used for re-organising entities consisting of single or multiple fields. 
The objectives of sorting can be classified into the following:

\begin{itemize}
\item \textit{Ordering} - arranging entities of the same type, or class into some ordered sequence.
\item \textit{Categorizing} - grouping or labelling entities with similar properties through sorting.
\end{itemize}

\begin{figure*}[t]
	\centering
	\includegraphics[trim = 0mm 108mm 0mm 5mm, width = 160mm]{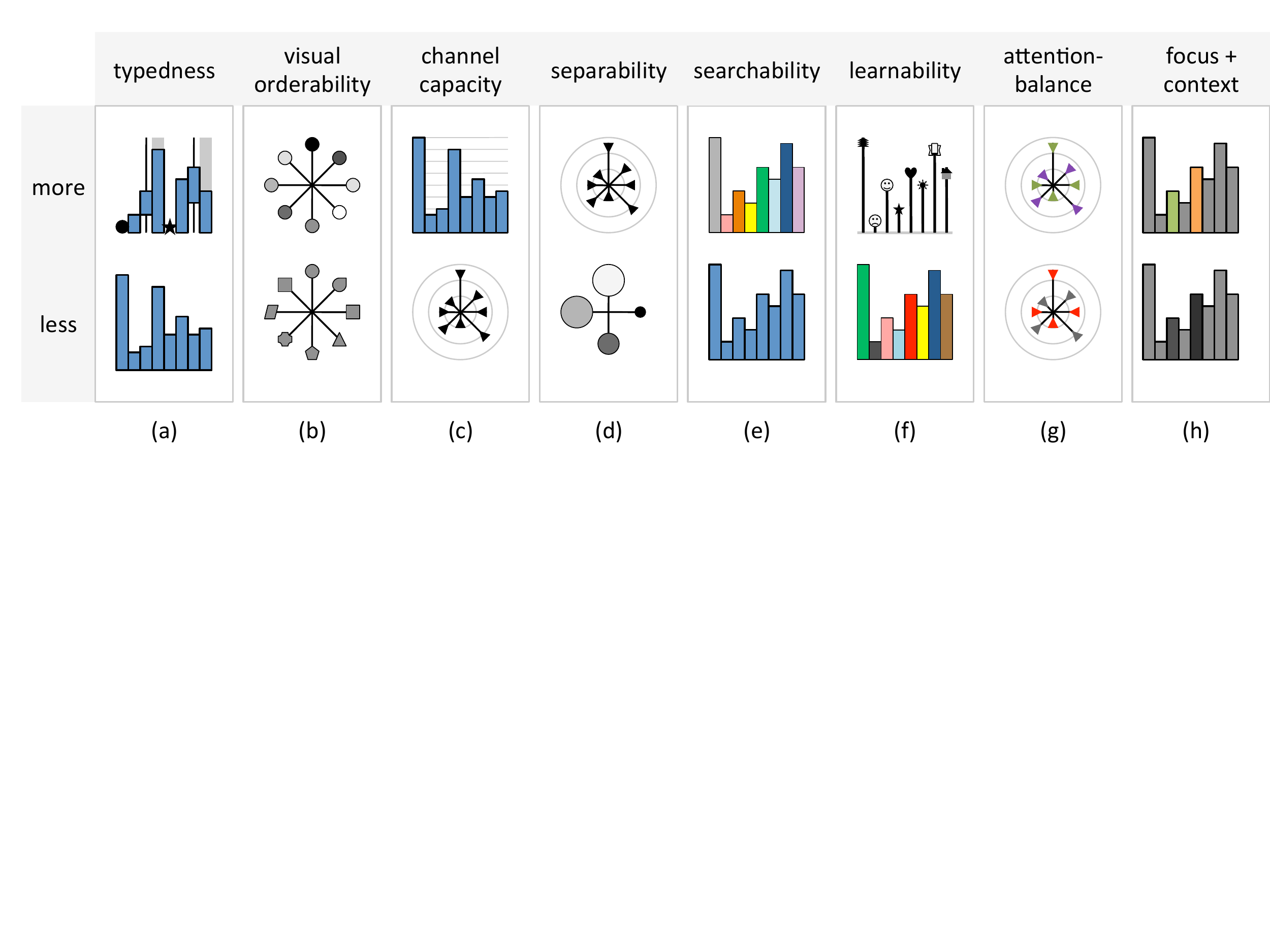}
	\caption{Variations of glyph design in accordance to the design principles of sortable glyph (a)-(h). For each principle, the top row depicts a glyph with greater emphasis and the bottom row depicts a glyph with less emphasis.}
	\label{fig:designprinciples}
\vspace*{-6mm}
\end{figure*}

A sort operation can be performed based on one or more attributes. We describe such attributes as \textit{sort keys}. 
In more general form, let us consider the set of objects or entities $E = (e_1, e_2, \ldots, e_s)$, each containing a set of attribute keys $K = (k_1, k_2, \ldots, k_n)$. 
This defines a $n$-dimensional attribute space which governs the sorting process. 
Thus, $e_i$ is a $n$-tuple or contains a $n$-tuple (as $e_i$ may have additional information such as a video clip).  
For example, a group of entities $E$ may be classified as a pack of cards (52 entities) which is sortable by keys $K$, such as card type (e.g., spades, clubs, diamond, and hearts), colour (e.g., red or black) or by value (1-13).

In order theory, we can specify two types of ordering relations: a weak (non-strict) order denoted by "$\preceq$", or a strict ordering "$\prec$".
These two properties characterize the mathematical concept of linear ordering~\cite{knuth73}. 
Given a subset of keys $\kappa \in K$, the goal of sorting is to arrange the entities $e_i$ into an ordered set (a list) such that
$ e^{\kappa}_1 \prec e^{\kappa}_2 \prec \ldots \prec e^{\kappa}_s $.
At the level of abstraction, sort keys as attributes can not be directly compared (i.e., by arithmetic =, and $<$, $>$), as they are essentially concepts. Hence, we introduce the notion $f^{\kappa} : E \mapsto \mathbb{R}$, that maps the object space with context keys $\kappa$ to a real value such that for any entity pair, $e_i, e_j$, the ordering relation $e^{\kappa}_i \prec e^{\kappa}_j$ implies:

\vspace*{-4mm}
$$ f^{\kappa}(e_i) < f^{\kappa}(e_j) \quad \forall i,j = 1, 2, \ldots, n. \quad i \neq j $$

With additional semantics, one can define such a function $f^\kappa$ to sort data (e.g., events) into more practical, or memorable orderings beyond common sorts (e.g., alphabetical), since $f^\kappa$ could be an importance function.
However, this may cause data to lose its perceived ordering at the analytical level.
We introduce glyph sorting as one solution for performing interactive sorting in visualization, where one goal is to use glyphs to sort the data.

\section{Design Principles of Sortable Glyphs}
\label{sec:designprinciples}

Building on previous works \cite{bertin83semiology, ward08, maguire12glyphTaxonomy}, we propose the following design principles for the creation of sortable glyphs to be used in interactive sorting as part of a visualization process.

\vspace*{-1mm}
\paragraph*{Typedness:}
Each variable in a multivariate dataset may be of a different data type. 
Typically, these are classified using the theory of scales~\cite{stevens46theoryOfScales} by: \emph{nominal}, \emph{ordinal}, \emph{interval}, and \emph{ratio}.
In addition, \emph{direction} should be considered as an important data type in visualization~\cite{ware08visualThinking}. 
Although hypothetically, we can map all data types to one or a few visual channels, such as length and size, it is more appropriate to use visual mappings that intuitively convey the underlying data type.
For example, in Figure~\ref{fig:designprinciples}(a) it is clearer to determine the underlying data types for each variable in the glyph from the top row (that illustrates greater emphasis) than the bottom row (that illustrates less emphasis).

\vspace*{-1mm}
\paragraph*{Visual Orderability:}
Some channels (e.g., size, greyscale intensity) naturally correspond to quantitative measures that enable a viewer to order different glyphs perceptually, while some others (e.g., an arbitrary set of shapes, or textures) are much more difficult for viewers to establish a consistent rule of ordering~\cite{ware08visualThinking, ware04perceptiondesign}.
Figure~\ref{fig:designprinciples}(b) shows two example glyphs depicting 8 variables of the same data type.
It is easier to visually order the 8 variables in the top glyph, than the bottom glyph.
Additional semantics can be attached to a visual channel such that it becomes visually orderable.
For instance, scientists often make use of the colour spectrum to determine the order of colours, which may not be natural to a child who is unfamiliar with this concept.
In some cases, one may have to use a visual channel with very poor orderability such as metaphoric pictograms.
The problem can be alleviated by accompanying such visual channels with an additional channel that is more visually orderable. 
For example, different pictograms can be associated with a background of different greyscales, or a regular polygonal boundary with different number of edges.
Alternatively, one may carefully design the pictogram set to make some components of pictograms orderable.
For example, Maguire \textit{et al.} designs a set of 7 pictograms with incremental number of components to encode levels of material granularity in biology \cite{maguire12glyphTaxonomy}.

\vspace*{-1mm}
\paragraph*{Channel Capacity:}
We adopt this term from information theory to indicate the number of values that may be encoded by a visual channel.
It is necessary to note that such a capability value is not an absolute quality, as the number depends on the size of a glyph as well as many other perceptual factors such as just noticeable difference~\cite{booth1993}, interference from nearby visual objects, or from a co-channel in an integrated channel~\cite{shepard64attention, handel72stimuli}.
From the glyph designs in Figure~\ref{fig:designprinciples}(c), we can clearly observe that the top glyph has a higher channel capacity since each bar can encode more values visually than the radial lines below.
It will always be desirable to use a visual channel with a higher capacity, though this is often in conflict with other requirements.

\vspace*{-1mm}
\paragraph*{Separability:}
There have been many psychology studies on the relative merits of separable and integrated visual channels (e.g., \cite{shepard64attention, handel72stimuli}).
Maguire \textit{et al.} discuss this requirement in the context of glyph design in \cite{maguire12glyphTaxonomy}.
We find that this requirement is particularly important to glyph sorting.
For example, in Figure~\ref{fig:designprinciples}(d), the glyph below encodes 8 variables using 4 integrated channels.
Each of the 4 circles encodes two variables using size and greyscale intensity.
Not only is the perception of individual channel affected by another in an integrated encoding, but also their ordering may demand more cognitive load in order for a viewer to detach one channel from another (e.g., intensity and size).

\vspace*{-1mm}
\paragraph*{Searchability:}
For glyphs encoding high-dimensional multivariate data, it is necessary to help viewers to search rapidly for a specific variable among many other variables~\cite{ware08visualThinking}.
In Figure~\ref{fig:designprinciples}(e), for example, it will be much easy to search for a green \emph{variable} than the 5th \emph{variable}.
Searchability is affected by many factors~\cite{healey12}. One dominant factor is the visual dissimilarity of individual channels. Hence searchability is closely related to \emph{typedness} and \emph{separability} as mentioned above.
It is also related to the spatial organisation of different visual channels such as grouping and ordering, as well as design appearance of each visual channel.
In many cases, one has to introduce an additional visual channel, such as colour in the top glyph in Figure~\ref{fig:designprinciples}(e) to help differentiate different variables.
Another factor is learnability, which is to be discussed below.

\vspace*{-1mm}
\paragraph*{Learnability:}
While legends are usually essential to glyph-based visualization systems, they cannot replace the need for careful glyph designs to help viewers learn and memorise the association between variables and visual channels without constantly consulting legends.
It is desirable for the appearance of a visual channel to be metaphorically associated with the semantic meaning of the corresponding variable~\cite{ware08visualThinking, sayim05}. 
One of the most effective metaphoric designs is to use pictograms. This design principle was demonstrated by Legg~\text{et al.}~\cite{legg12} through the deployment of glyph-based visualization in sports.
Figure~\ref{fig:designprinciples}(f) shows two different levels of learnability, when for example one needs to encode the number of greeting cards in different categories.
The glyph on the top row is semantically rich and is much easier to learn than that on the bottom row.
However, not all glyph-based visualization can afford pictograms.
These constraints can often be alleviated by making abstract metaphoric association, such as green for nature, renewable, safe, and so on.

\vspace*{-1mm}
\paragraph*{Attention Balance:}
In multivariate visualization, one common task is to make observation of the ``\emph{behaviour}'' of different variables in relation to the variable(s) in a sorted order.
While it is helpful to make each individual variable searchable~\cite{tsot95, ware08visualThinking}, it is also necessary to avoid unbalanced attentiveness among different channels.
For example, the bottom glyph in Figure~\ref{fig:designprinciples}(g) features bright red indicators for some variables.
When browsing different glyphs in visualization, these red triangles are dominant which may cause undesirable pop-out effects. 

\vspace*{-1mm}
\paragraph*{Focus + Context:}
In multivariate visualization, it is usually difficult, often undesirable, to pre-determine what is the focus variable and what is the context variable.
Naturally, in glyph sorting, a variable that is associated with a sort key is considered as one of the foci.
In some cases, the viewer may wish to consider another variable as a focus.
Hence it is desirable for a glyph sorting system to support focus+context visualization by highlighting individual channels that are in focus.
Straka~\textit{et al.}~\cite{straka04vesselGlyph} demonstrates this design principle for glyphs in CT-angiography.
This can be expensive, because in the worst case, each visual channel is accompanied with another channel as a highlighter.

\vspace*{-1mm}
\paragraph*{Labelling and Legends:}
Axis-labelling is an essential requirement for any sorting configuration for indicating sort keys~\cite{ward10interactive}. It enables the viewer to understand the context (e.g., frequency vs. amplitude in sound analysis) without referring to the visualization itself. Bertin~\cite{bertin83semiology} refers to this as external identification.
Legends convey the relationships between variables and visual channels and its representation for a given discrete or continuous value.
This is often known as internal identification~\cite{bertin83semiology}. 

These design principles are general guidelines that we consider when designing glyphs to be sorted interactively in visualization.
However, they should not be treated as the absolute laws.
Some cases may lead to conflicting requirements when following some of these principles, or compete for limited capacity of visual channels for smaller designs.
\vspace*{-2mm}
\section{Interactive, Glyph-based Visual System}
\label{sec:interactivesystem}

\begin{figure}[t]
\centering
\includegraphics[trim = 0mm 35mm 100mm 0mm, width = 80mm]{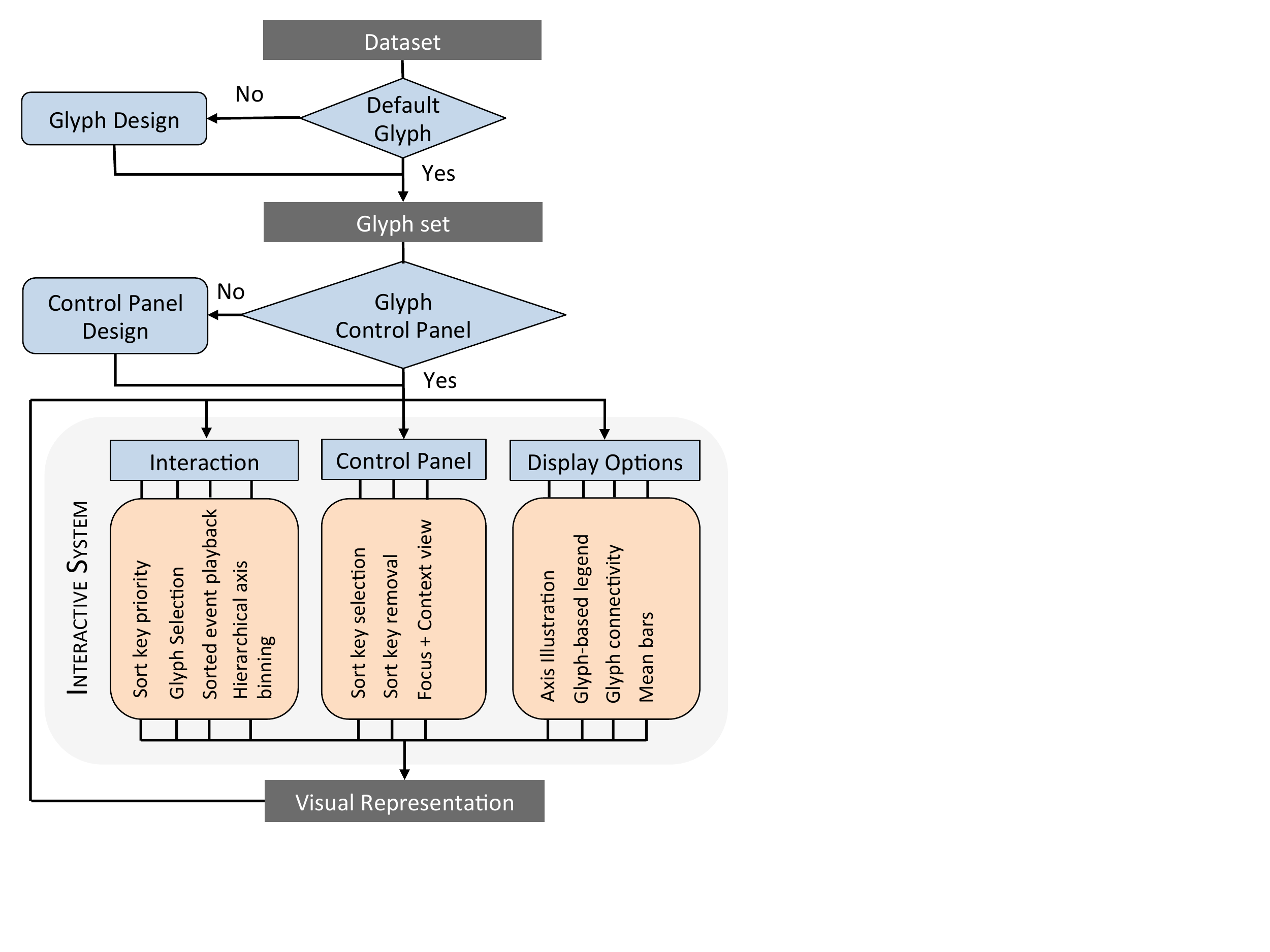}
\caption{A graphical pipeline illustrating the glyph sorting framework. It consists of four key steps: 1) visual mapping of data to glyphs. 
We propose general design guidelines for creating visually sortable glyphs to support interactive sorting and multivariate analysis.
Alternatively, a default glyph (e.g., Star glyph) is used. 2) integrating a focus and context glyph control panel for selecting multiple sort keys, 3) constructing the glyph sorting tool which enables users to perform high-dimensional sorting and interactively adjust various display options and 4) visual representation of sorted results on a Interactive, Multi-dimensional glyph plot.}
\label{fig:sortingpipeline}
\vspace*{-6mm}
\end{figure}

In this section, we propose a glyph-based visual analytic system for performing glyph sorting which is outlined in Figure~\ref{fig:sortingpipeline}. 
The system integrates two fundamental components: 1) a glyph control panel for selecting and driving the sorting process in a visual manner, and 2) an Interactive, Multi-dimensional Glyph plot for viewing sorted results.

\vspace*{-2mm}
\subsection{Focus and Context Glyph-based Interface}
\label{sec:interface}

Our glyph-based, sorting system utilises a focus and context glyph-based user-interface for selecting sort keys [see supplementary video]. 
The interface provides two main benefits. It allows users to interactively control the sorting process by populating sort keys within the linked IMG plot in a visually intuitive manner. Secondly, the focus and context glyph gives a visual reference which allows users to rapidly identify and understand the attributes that drive the sorting.

Sort keys are selected in the system by interactively clicking on a visual component of the glyph. The selected visual attribute is then rendered into focus using opacity such that the data attribute is visually distinct from other attributes. This is an effective method for emphasising specific parts to the users attention in high-dimensional glyphs.   
Similarly, users can remove a sort key by clicking on a glyph component in focus and dragging it off the glyph to bring the attribute back into context.
By linking the interface with the IMG plot, users are able to populate different sort keys in a visually intuitive manner.
Furthermore, we incorporate tooltips into the interface to aid users with information on what attributes is visually encoded in each glyph component.

\vspace*{-2mm}
\subsection{Interactive, Multi-dimensional Glyph Plot}
\label{sec:IMGplot}

Since ordering in a sorting plane is one of the most effective and widely recognised representations for data analysis (e.g., scatter plot), we position the glyphs along the two primary sorting axes. This forms the basis of our Interactive, Multidimensional Glyph (IMG) plot.
Following the design principles in Section~\ref{sec:designprinciples}, populated sort keys are depicted as focus and context glyphs along each sorting axis respectively (see Figure~\ref{fig:matchcomparison} for example) coupled with a visual legend to illustrate how the data is ordered. The sort key priority can be changed interactively by the user via double clicking on the sort key glyph, to either promote (using the left mouse button) or demote (using the right mouse button) the ordering.
We integrate a series of interactive tools to aid user exploration: sliders for adjusting axis length, brushing tools for selecting glyphs, pan-and-zoom navigation for details on demand and viewing of additional information (e.g., a video, or image) that may be associated with a glyph.

Visualizing glyphs on a 2D plane imposes additional challenges. 
One perceptual problem is the order in which glyphs are rendered on the IMG plot. By default, glyphs are rendered sequentially as they occur in the dataset.
Depending on the sorting parameters, these will cause different levels of overlap.
To alleviate this, we incorporate the ability to sort the rendering order of selected glyphs.
This enables the user to emphasise glyphs of greater interest for data exploration.
In addition, we provide two display preferences as a user-option. \emph{Connectivity}, for rendering lines that connect glyphs in order of a sorting attribute, and \emph{Mean Bars} which displays the statistical average value of a sorting axis (if applicable) as a coloured band in each hierarchical axis bin.

\subsection{Hierarchical Axis Binning}
\label{sec:axisbinning}

In data-driven placement, sorting data by discrete variables is a typical operation one can perform. 
However, this often leads to an increase in level of overlap due to discrete positioning in the constrained sorting space.
Ward~\cite{ward02} describes a survey on distortion techniques (e.g., random jitter~\cite{cleveland93}), a post-processing step which can be used to reduce visual clutter by incorporating noise into the glyphs position.
A major concern with this approach is the level of distortion introduced can significantly change the interpretation and integrity of the visualization.

Hierarchical axis binning is a mapping function to alleviate such a problem by representing multiple discrete variables as regions as opposed to points.
Encoding multiple dimensions onto a single axis enables additional sorting functions (e.g., a continuous variable) to be mapped for relaxing the positioning of glyphs along a bounded sub-region.
Figure~\ref{fig:sortingblocks} illustrates our generalised axis binning algorithm at different levels of sorting which we demonstrate along one axis. However, our technique can be applied over multiple sorting axes.
Let $L$ be the interval $[L_{\min}, L_{\max}]$ and $K = (k_1, k_2, \ldots, k_n)$ be a set of sort keys we want to order the data by. We define our axis mapping function for a single key $k$ as the following:

\vspace*{-4mm}
\begin{equation}
h(e, L, k) = \frac{f^{k}(e)}{\max f^{k}} || L || + L_{\min}
\label{eq:axisMapping}
\end{equation}

The linear function first normalises the attribute key and maps this to the region $L$, such that if $k$ is discrete and non-numerical (e.g., name), then $\max f^{k}$ is equivalent to the cardinality $||k||$ of the sort key.
For higher order sorting, we expand the region given by each discrete value hierarchically to map additional sort functions. 
Let us first denote the type of a key as $k^T$, where $T =~\{Discrete, Continuous\}$.
Now suppose $A \in K$ is a ordered sequence of discrete and continuous sort keys. 
We apply the restriction $A^{T_i}_i = \{ a^{T_1}_1, \ldots, a^{T_n}_n \}$ such that $T_i \preccurlyeq T_{i+1}$ for $i = 1, \ldots, n-1$, where the condition $\preccurlyeq$ is used to obtain a list where no continuous key directly precedes a discrete key for each sort key pair.
With such an ordered list, we can define a hierarchical sorting function for mapping and relaxing points along discrete sub-regions recursively by Eq.~\ref{eq:axisMapping}.
This is generalised to the following form:

\begin{figure}[t]
\centering
\includegraphics[width = 7.5cm]{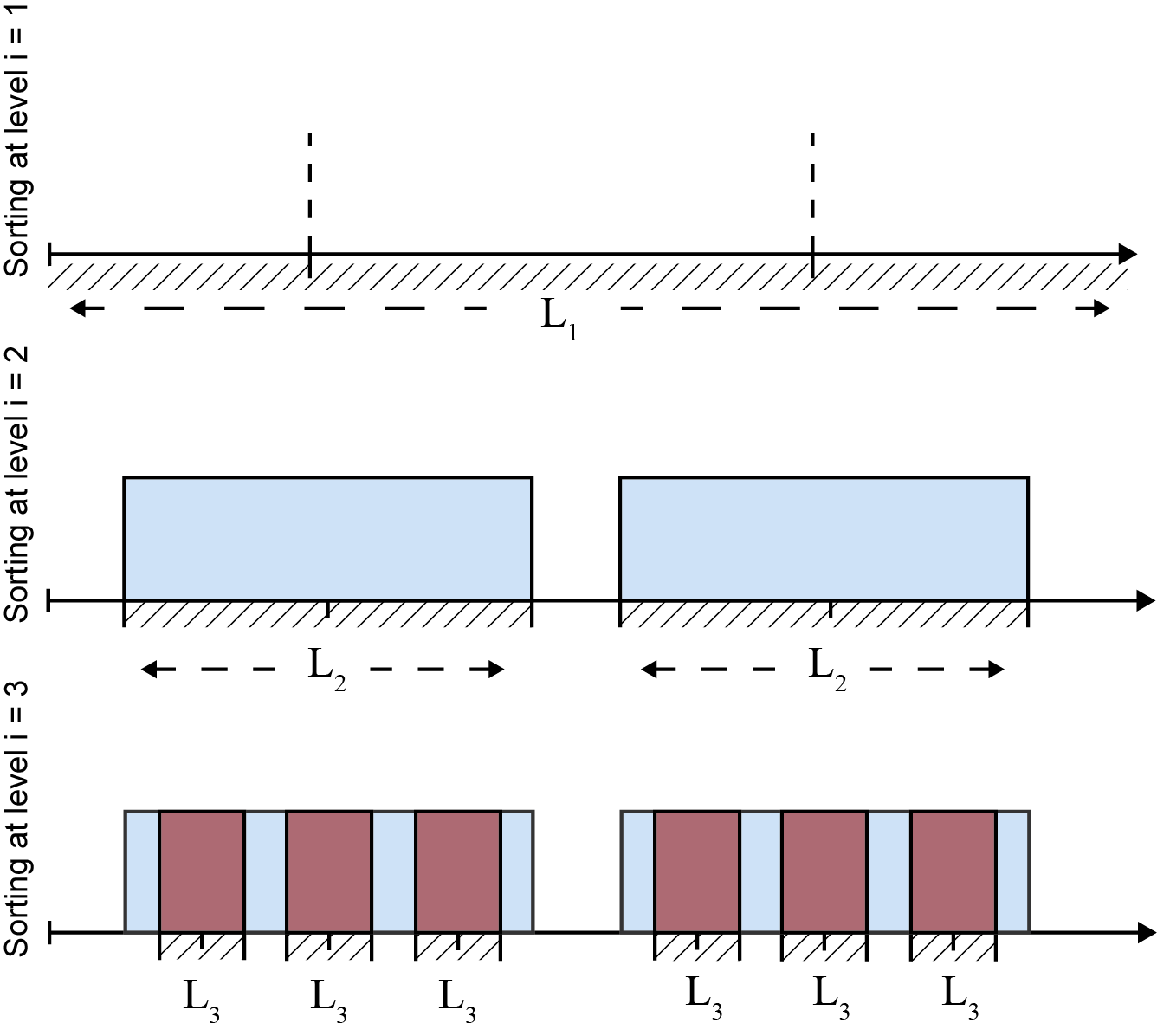}
\vspace*{-3mm}
\caption{Diagram illustrating hierarchical axis binning along one sorting axis. Intervals (or sub-regions) at each level of sorting can be sub-divided by different attributes where additional sort functions can be mapped. We note that the axis mapping can be applied along multiple axes.}
\label{fig:sortingblocks}
\vspace*{-6mm}
\end{figure}

\vspace*{-4mm}
\begin{equation}
H(e,L,A) = \sum_{i = 1}^{n} h (e, L_i, a_i)
\end{equation}

where $L_{i}$ is the interval at each level. At the sort level $i=1$, our interval is already initialised (i.e., the axis length, where $L_1 = L$). Thus, it is only necessary to determine each sub-region division at successive levels of sorting. The sub regions are defined as $L_{i+1} \in [-\delta_{i+1}, +\delta_{i+1}]$ such that:

\vspace*{-4mm}
\begin{equation}
\delta_{i+1} = \frac{|L_i|}{2 \max f^k} \cdot \mu, \quad\quad \mu \in [0,1)
\end{equation}

in which the coefficient $\mu$ is used to adjust the maximum length of each sub region. For $\mu = 1$, adjacent sub regions touch (connected), while for $\mu > 1$, our intervals begin to overlap. For visual representation, we set $\mu \leq 0.8$ to allow significant gaps between each axis bin at all levels of sorting.  
Since our function is bijective, it follows that each data point is unique.
Hence, the complexity of ordering glyphs with multiple sort keys both analytically and visually is reduced to sorting by a dominance relation (e.g., x and y coordinate).

Given a ordered list of discrete and continuous keys, we can hierarchically build multiple axis bins to facilitate sorting of multiple functions.
The user is able to interactively control the amount of spatial relaxation by adjusting two properties: the axis length and the width of axis binning. 
Each hierarchical axis bin size is altered by varying the sorting parameter $\mu$ which corresponds to each level of sorting.

\section{Case Study: Sports Event Analysis}
\label{sec:evaluation}
We demonstrate glyph sorting on a real-world application in sports event analysis. We have worked in close collaboration with the Welsh Rugby Union (WRU) to develop a software that allows for in-depth analysis of matches. First, we detail the process of mapping attributes to a sortable glyph. We then present a visual comparison of two matches which was conducted by analysts at the WRU. We discuss the knowledge and insight that has been derived as a result of glyph sorting and conclude the study with domain expert feedback.

\subsection{Visual Mapping of Sort Keys}
\label{sec:visualmapping}

\begin{figure}[t]
\centering
\includegraphics[trim = 10mm 120mm 60mm 5mm, width = 80mm]{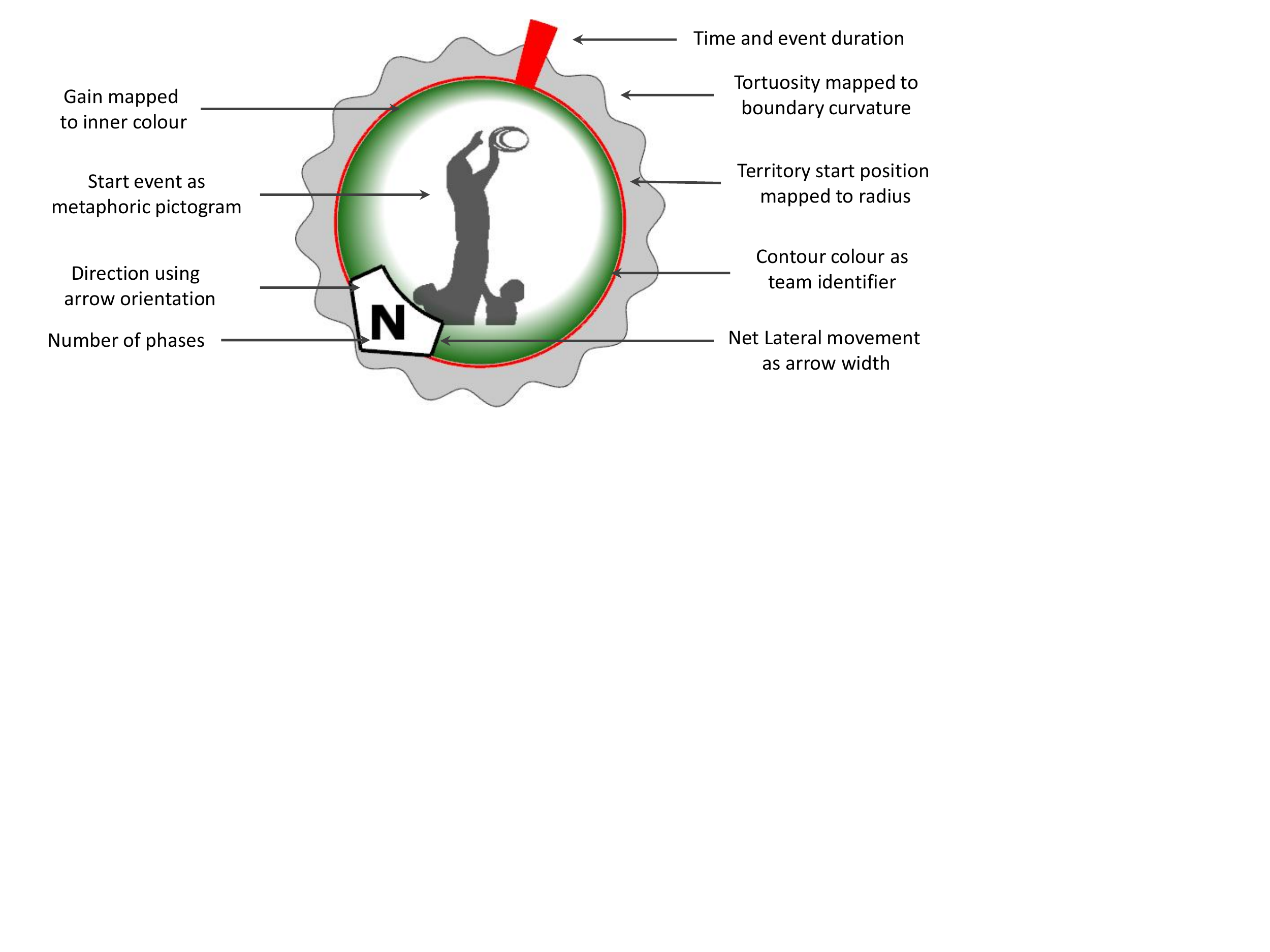}
\caption{Components and visual channels of the glyph.}
\label{fig:finalglyph}
\vspace*{-6mm}
\end{figure}

In sports performance analysis, coaches and analysts heavily rely on notational data~\cite{hughes97}. This involves \textit{"tagging"} video footage with key events and semantic notations from which key performance indicators can be derived.
Spatial tracking data is another source of information which analysts study as a separate field. However, without the semantic context, such data is meaningless and is often disregarded due to the deluge of data.
We design glyphs that combine both notation and spatial data to be used for interactive sorting and visualization [see supplementary Figure A].
Table~\ref{table:sortattributes} gives an overview of the set of attributes in rugby event analysis which are ranked in order of data importance based on end-user feedback.
Following the design principles presented in Section~\ref{sec:designprinciples}, we describe the methodology of mapping rugby event data to visually sortable glyphs (see Figure~\ref{fig:finalglyph}).  


\begin{table}[t]
 \begin{center}
   \begin{tabular}{cccc}
     Sort Key & Typedness & Visual Channel\\
   \hline
      Gain & Ordinal & Colour\\ 
      Event & Nominal & Pictogram\\
      Territory Start Position & Interval & Size \\          	
      Tortuosity & Ratio & Shape \\
      Number of Phases & Ratio & Enumerate\\
      Direction & Direction & Orientation \\      
      Net Lateral Movement & Ratio & Length\\ 
      Time & Ratio & Location \\
      Phase Duration & Ratio & Length\\
      Team Identifier & Nominal & Colour \\  
   \end{tabular}
 \end{center}
  \caption{Table illustrating the set of sort keys in rugby event analysis. Each attribute is classified based on typedness, and the visual channel mapped to the glyph. Data attributes are ranked in order of importance from top to bottom.}
   \label{table:sortattributes}
   \vspace*{-6mm}
\end{table}

The goal of rugby is to carry a ball to the opposition try line.
\emph{Gain} is the term used for the distance gained towards the opposition try line as a result of free play. Although gain is naturally of interval type, conventions in rugby adopt an ordinal measurement (e.g., negative gain, minor variation, major gain). Thus, a discrete representation is needed. 
Since end-users make use of an existing ordered colour scheme, it is natural to map gain to this visual channel to support visual orderability, learnability as well as being searchable given the high visual priority of colour.
The context in which gain is achieved is particularly important. These \emph{start events} (e.g., from lineout, scrum, etc.) are nominal categories that classifies periods of play into more semantically meaningful groups. Here, the events are sorted by importance.
We discuss previously in Section~\ref{sec:designprinciples} the use of metaphoric pictograms for mapping such data~\cite{legg12}. 
Pictograms can often be arbitrary, in that their shape, size, colour will vary, thus having a low visual orderability. Using different intensities to draw each pictogram is one solution to establish a visual ordering, however, this may be misleading since event is discrete and not continuous. 
Instead, we design and order the pictograms according to their relative greyscale pixel count which is more appropriate for our study.
Typically associated with a start event, is whether that event resulted in point scored (i.e., the end event).
These glyphs should be differentiable to the viewer. Therefore, we use a coloured halo effect to enhance the attention-balance of such glyphs.

In rugby, the pitch is divided into key areas known as territory, which describes the spatial property of an event.
The \emph{territory start position} gives an indication at how far an event occurs from the opposition try line.
Given that visual separability of variables is a key requirement in glyph sorting, we avoid overloading a single channel (e.g., colour) by encoding this attribute using size.
Using the glyph template described in~\cite{maguire12glyphTaxonomy}, we map this to the radius of a transparent, external grey silhouette. 
Size is a suitable mapping for ordering quantitative variables (i.e., interval and ratio) and also yields a high searchability due to visual pop-out, making this ideal for attributes of greater importance.
The additional channel capacity introduced by the silhouette enables us to encode a varying line curvature along the contour for displaying the \emph{tortuosity} of the ball path. 
Semantically, the line curvature resembles the tortuosity or shape of the ball path, which makes this easier for users to infer or remember.

A single path (or ball-in-phase), consists of a series of waypoints and path segments. In rugby, these waypoints and segments correspond to the \emph{number of phases}. A simple and effective mapping for such discrete data is to use a enumerative representation due to its natural ordering. 
We depict the enumerate inside an arrow head which is oriented according to the resulting ball \emph{direction}.
Since orientation has weak learnability, we incorporate metaphoric cues i.e., a compass, by positioning the arrow head along a circle to make this more memorable to the end-user. We map arrow width to \emph{net lateral movement} which indicates the relative lateral distance travelled. Since net lateral movement and direction is co-related, it is sensible to couple both variables together.

\begin{figure*}
	\centering
	\includegraphics[width = 17cm]{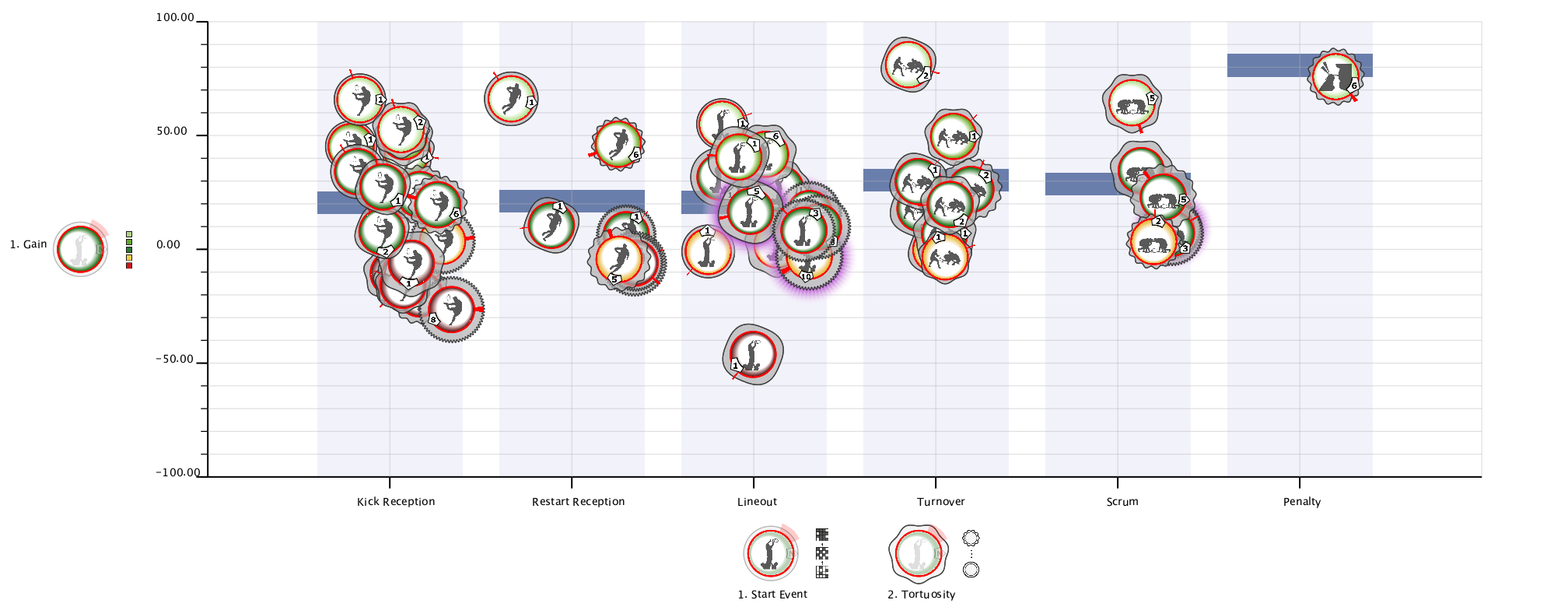}
	\includegraphics[width = 17cm]{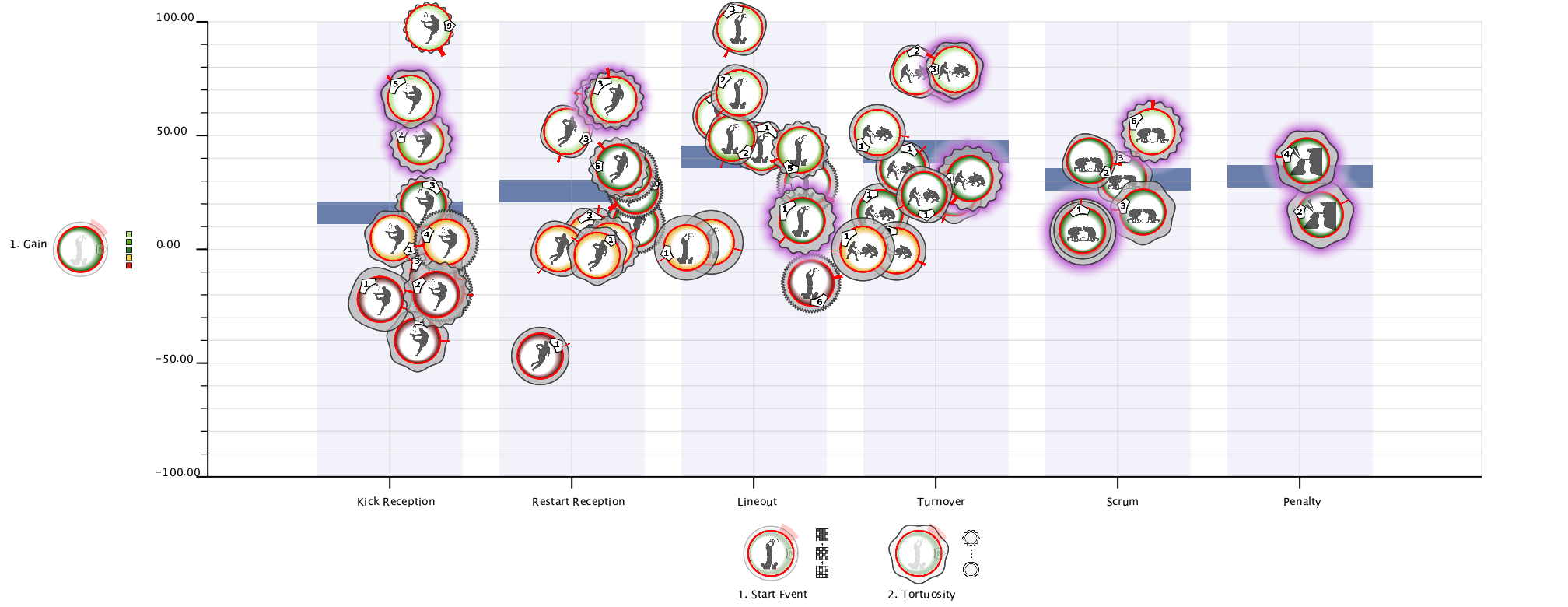}
	\vspace*{-6mm}
	\caption{Visual comparison of two rugby matches using the IMG plot. Top: Match 1 (\textbf{M1}). Bottom: Match 2 (\textbf{M2}). For each match, the glyphs are ordered using three sort keys: Gain versus Start Event and Tortuosity. Mean bars are also displayed as a user-option to provide additional statistical information.}
\label{fig:matchcomparison}
\vspace*{-6mm}
\end{figure*}

Another data coupling is \emph{time} and \emph{phase duration} which describes the temporal period in which the event occurs. 
Because both attributes are of ratio type and continuous, it is possible to combine such data using an integrated encoding, for maximising channel capacity. We represent time using a clock visual metaphor, where time and duration is mapped to location (or orientation) and length of the time handle. 
The semantics of a clock is used to enhance the visual orderability property of time.
In order to facilitate aspects of our sort key visual mappings, we adopt a circular-based glyph design (Figure~\ref{fig:finalglyph}). 
The final attribute we map is \emph{team identifier} (i.e., home or opposition), which we depict by colour-coding the inner contour.
We follow the general convention used in sport for distinguishing two teams by mapping red and blue to the teams respectively.
This enables sporting domain experts to be more familiar with the glyph concept which relates to learnability and visual search.

\subsection{Visual Comparison of Two Matches}
\label{sec:visualcomparison}

Analysts are normally tasked with watching multiple match videos to identify the occurrences of key performances. This is laborious and time-consuming, and even current techniques such as notational analysis do not allow the analysts to discover new insight but merely review what has been previously recorded. As part of our evaluation, we compare glyph-based visual analytics for analysing the performance of a single team in two different rugby matches as shown in Figure~\ref{fig:matchcomparison}. 
Match 1 (\textbf{M1}), involves two evenly matched teams, resulting in a closer point score differential.
This is compared to Match 2 (\textbf{M2}) where one team proved to be more dominant.
Both matches are taken from the World Cup 2011.
By using visual analytics, the domain experts are interested to see how the two matches compare and for investigating why the outcome of the two matches are so different.

We presented the software to the analysts and explained the usability prior to letting the analysts explore the two datasets.
One topic of interest is the relationship between gain and tortuosity, i.e., whether the strategy of working the opposition (high tortuosity) resulted in greater gain. 
Sorting the glyphs by the two attributes reveals a uniform gaussian distribution of glyphs in both matches [see supplementary video].
A clear observation, is the significantly lower average tortuosity in \textbf{M2}, indicated by the greater spread of glyphs and overall shift along the tortuosity sorting axis.
This shows that it requires less effort to make sizeable gains in each phase and thus, attacking the opposition more directly can yield larger benefits.
From a glance, the analyst identifies many different event types (i.e., pictograms) appearing within the cluster in the visualization. This directed the user to inspect how start event would affect the ordering of the glyphs via an alternative sorting strategy.

Figure~\ref{fig:matchcomparison} illustrates the comparison of the two matches where the user sorts the glyphs based on three attributes: gain versus start event and tortuosity.
One new feature not previously observable, is the variation of start events that resulted in points scored which is depicted by glyphs highlighted in purple.
It is clear in \textbf{M1}, that most points are scored from lineouts. In comparison, the other match exhibits a more uniform distribution of point scoring events.
From this, we can hypothesise specific strengths and weakness of different teams.
The statistical information displayed by mean bars is useful for analysing and deriving new key performance indicators. For instance, phases from turnover provide the most average gain across both matches as shown by the highest blue bands in each hierarchical axis bin. 
Thus, the number of turnovers is one key indicator that influences the team performance. 
Subsequently, scrums is the next most effective in \textbf{M1}, whereas lineouts proved more successful in \textbf{M2}.

Under the new sort operation, the analysts discover a new data trend that is present in \textbf{M1} and not the other (see Figure~\ref{fig:matchcomparison}), where the glyphs appear within each axis bin along a linear line from top left to bottom right.
This indicates that the team achieved more gain whilst attacking the opposition directly, which decreases respectively with higher tortuosity.
At first, this was not what the analysts expected.
By visually analysing the glyphs in the upper left cluster, we found the events to occur largely within the defensive third as shown by the shorter grey silhouette on each glyph.
For a greater level of detail, the analyst studies the sorted video clips that is associated with each glyph, to find the cause of higher gain is a result to the team kicking the ball forward out of defense.
Although kicking the ball results in greater gain, this comes at a cost of losing ball possession which is crucial.

The analysts found the trends to be insightful for explaining strategies against different oppositions.
Tactically, the visual patterns observed in \textbf{M2} describes a more offensive game plan which is carried out each time the team regained ball possession.
In comparison, \textbf{M1} shows a clear distinction between offense and defense, where the team selectively chose key moments (e.g., pitch position) to attack the opposition.
The information correlates well with the analysts understanding since mistakes against stronger oppositions (i.e., \textbf{M1}) comes with higher risk which can impact the outcome of a match.
One further observation visible in \textbf{M2} is shown by the ordering of glyphs in the turnover event category, in which the variable gain increases with tortuosity.
Such a pattern indicates the opposition defence tiring as the home team attacked the ball, creating a prospective scoring opportunity.

\vspace*{-2mm}
\subsection{Domain Expert Review}
\label{sec:expertreview}
The development of the work has been an iterative process in close collaboration with the Welsh Rugby Union (WRU), spanning over 12 months. From inception of the idea, it was clear that the analysts want to be able to interrogate their data in a more complex nature than previously available in order to gain new insight. The introduction of spatial data into visual analytics has meant that this is now achievable and has been used to derive novel information intuitively. [name remove for review] of the WRU performance analysis team provide valuable feedback on the usage of glyph sorting within rugby performance analysis.

\textit{``The strongest element of the system is the ability to interactively sort vast quantities of data according to multiple attributes for revealing trends or groups of data.
Your eyes are instantly drawn to those patterns.
In our current practice, getting the data and generating charts (through spreadsheets) is very time consuming. Once a chart is plotted, we often get "What if we take this variable into account?", which then requires us to go back to the raw data and process it all again. Where as with this, we can navigate the data much more effectively.
The visualization is insightful for giving an overview of a match. Sorting the data gives good visual cues for pointing us in the right direction and being able to look in detail at the associated videos helps to clarify and explain what those trends are.''}

The feedback received from the WRU analysis team proved to be very encouraging. It confirms that the use of glyph sorting can significantly enhance the effectiveness of glyph-based visualization.
By integrating glyphs into the sorting process and linking this with multiple video footage, the analyst is able to derive new underlying phenomena from a match.
In particular, the domain expert feel that such a system is highly beneficial in their workflow for post-match analysis, where the insight obtained from sorting is useful for formulating strategies against different oppositions.


\vspace*{-2mm}
\section{Conclusions}
\label{sec:conclusion}

In this work we have developed a glyph-based sorting framework for interrogating and interpreting large multivariate data. We have demonstrated the technique by applying it to sports performance analysis, where a variety of continuous and discrete data forms are incorporated into a visually sortable glyph design.
Glyph sorting is an effective means for multivariate analysis and can be used to enhance the usability of glyph-based visualization and enrich the users with alternative sorting strategies for revealing trends. 
Our sorting framework enables the analysts to derive new insight as a result of high-dimensional sorting that was previously not observable with existing techniques. 


\bibliographystyle{eg-alpha}
\bibliography{glyphliterature}

\newcommand{\etalchar}[1]{$^{#1}$}
\begin{thebibliography}{\uppercase{MPRSDC12}}

\bibitem[BBS{\etalchar{*}}08]{botchen08ActionBasedVideo}
\textsc{Botchen R.~P., Bachthaler S., Schick F., Chen M., Mori G., Weiskopf D.,
  Ertl T.}:
\newblock Action-based multifield video visualization.
\newblock \emph{IEEE Transactions on Visualization and Computer Graphics 14}, 4
  (2008), 885--899.

\bibitem[Ber83]{bertin83semiology}
\textsc{Bertin J.}:
\newblock \emph{Semiology of graphics}.
\newblock University of Wisconsin Press, 1983.

\bibitem[BF93]{booth1993}
\textsc{Booth D.~A., Freeman R. P.~J.}:
\newblock Discrimnative measurement of feature integration in object
  recognition.
\newblock \emph{Acta Psychologica 84} (1993), 1--16.

\bibitem[BFCM06]{bender06}
\textsc{Bender M.~A., Farach-Colton M., Mosteiro M.}:
\newblock Insertion sort is {O}(n log n).
\newblock \emph{Theory of Computing Systems 39}, 3 (2006), 391--397.

\bibitem[Che73]{chernoff73}
\textsc{Chernoff H.}:
\newblock Using faces to represent points in $k$-dimensional space graphically.
\newblock \emph{Journal of the American Statistical Association 68} (1973),
  361--368.

\bibitem[Cle93]{cleveland93}
\textsc{Cleveland W.~S.}:
\newblock \emph{Visualizing Data}.
\newblock Hobart Press: Summit, 1993.

\bibitem[CR05]{chlan05}
\textsc{Chlan E.~B., Rheingans P.}:
\newblock Multivariate glyphs for multi-object clusters.
\newblock In \emph{Information Visualization, 2005. INFOVIS 2005. IEEE
  Symposium on} (2005), pp.~141--148.

\bibitem[Dem56]{demuth56}
\textsc{Demuth H.}:
\newblock \emph{Electronic Data Sorting}.
\newblock PhD thesis, 1956.

\bibitem[dLvW93]{deLeeuwVanWijk93probe}
\textsc{de~Leeuw W.~C., van Wijk J.~J.}:
\newblock A probe for local flow field visualization.
\newblock pp.~39--45.

\bibitem[ECW92]{estivill-castro92}
\textsc{Estivill-Castro V., Wood D.}:
\newblock A survey of adaptive sorting algorithms.
\newblock \emph{ACM Comput. Surv. 24}, 4 (Dec. 1992), 441--476.

\bibitem[HE12]{healey12}
\textsc{Healey C., Enns J.}:
\newblock Attention and visual memory in visualization and computer graphics.
\newblock \emph{IEEE Transactions on Visualization and Computer Graphics 18}, 7
  (2012), 1170--1188.

\bibitem[HF97]{hughes97}
\textsc{Hughes M.~D., Franks I.~M.}:
\newblock \emph{Notational analysis of sport}.
\newblock London: E. \& F.N. Spon., 1997.

\bibitem[HI72]{handel72stimuli}
\textsc{Handel S., Imai S.}:
\newblock The free classification of analyzable and unanalyzable stimuli.
\newblock \emph{Perception and Psychophysics} (1972), 108--116.

\bibitem[HLNW11]{hlawatsch11flowRadar}
\textsc{Hlawatsch M., Leube P., Nowak W., Weiskopf D.}:
\newblock Flow radar glyphs - static visualization of unsteady flow with
  uncertainty.
\newblock In \emph{VisWeek} (2011).

\bibitem[Hoa62]{hoare62}
\textsc{Hoare C. A.~R.}:
\newblock Quicksort.
\newblock \emph{The Computer Journal 5}, 1 (1962), 10--16.

\bibitem[JBMC10]{heike10soundRiver}
\textsc{J\"{a}nicke H., Borgo R., Mason J. S.~D., Chen M.}:
\newblock Soundriver: Semantically-rich sound illustration.
\newblock \emph{Computer Graphics Forum 29}, 2 (2010), 357--366.

\bibitem[Knu98]{knuth73}
\textsc{Knuth D.~E.}:
\newblock \emph{The Art of Computer Programming, Vol. 3: Sorting and Searching,
  Second Edition}.
\newblock Addison-Wesley, Reading, Mass., 1998.

\bibitem[LAK{\etalchar{*}}98]{laidlaw98mousespinal}
\textsc{Laidlaw D.~H., Ahrens E.~T., Kremers D., Avalos M.~J., Jacobs R.~E.,
  Readhead C.}:
\newblock Visualizing diffusion tensor images of the mouse spinal cord.
\newblock In \emph{IEEE Visualization} (1998), pp.~127--134.

\bibitem[LCP{\etalchar{*}}12]{legg12}
\textsc{Legg P.~A., Chung D. H.~S., Parry M.~L., Jones M.~W., Long R.,
  Griffiths I.~W., Chen M.}:
\newblock Matchpad: Interactive glyph-based visualization for real-time sports
  performance analysis.
\newblock \emph{Computer Graphics Forum 31}, 3pt4 (2012), 1255--1264.

\bibitem[LKH09]{lie09criticaldesign}
\textsc{Lie A.~E., Kehrer J., Hauser H.}:
\newblock Critical design and realization aspects of glyph-based 3d data
  visualization.
\newblock In \emph{Proceedings of the 2009 Spring Conference on Computer
  Graphics} (2009), SCCG '09, ACM, pp.~19--26.

\bibitem[MPRSDC12]{maguire12glyphTaxonomy}
\textsc{Maguire E., P.~Rocca-Serra S.-A.~S., Davies J., Chen M.}:
\newblock Taxonomy-based glyph design: with a case study on visualizing
  workflows of biological experiments.
\newblock \emph{IEEE Transactions on Visualization and Computer Graphics}
  (2012).

\bibitem[PRdJ07]{pearlman07}
\textsc{Pearlman J., Rheingans P., des Jardins M.}:
\newblock Visualizing diversity and depth over a set of objects.
\newblock \emph{IEEE Computer Graphics and Applications 27} (September 2007),
  35--45.

\bibitem[RP08]{ropinski08}
\textsc{Ropinski T., Preim B.}:
\newblock Taxonomy and usage guidelines for glyph-based medical visualization.
\newblock In \emph{{Proc. Simulation and Visualization}} (2008), pp.~121--138.

\bibitem[SCC{\etalchar{*}}04]{straka04vesselGlyph}
\textsc{Straka M., Cervenansky M., Cruz A.~L., Kochl A., Sramek M., Groller E.,
  Fleischmann D.}:
\newblock The vesselglyph: Focus \& context visualization in ct-angiography.
\newblock In \emph{Proceedings of the conference on Visualization} (2004),
  pp.~385--392.

\bibitem[SEK{\etalchar{*}}98]{shaw98}
\textsc{Shaw C., Ebert D., Kukla J., Zwa A., Soboroff I., Roberts D.}:
\newblock Data visualization using automatic, perceptually-motivated shapes.
\newblock In \emph{Visual Data Exploration and Analysis} (1998), SPIE.

\bibitem[SFGF72]{siegel72}
\textsc{Siegel J., Farrell E., Goldwyn R., Friedman H.}:
\newblock The surgical implication of physiologic patterns in myocardial
  infarction shock.
\newblock \emph{Surgery 72} (1972), 27--35.

\bibitem[She64]{shepard64attention}
\textsc{Shepard R.}:
\newblock Attention and the metric structure of the stimulus space.
\newblock \emph{Journal of Mathematical Psychology 1} (1964), 54--87.

\bibitem[SJAS05]{sayim05}
\textsc{Sayim B., Jameson K.~A., Alvarado N., Szeszel M.~K.}:
\newblock Semantic and perceptual representations of color: Evidence of a
  shared color-naming function.
\newblock \emph{The Journal of Cognition and Culture 5}, 3-4 (2005), 427--486.

\bibitem[Ste46]{stevens46theoryOfScales}
\textsc{Stevens S.~S.}:
\newblock On the theory of scales of measurement.
\newblock \emph{Science 103}, 2684 (1946), 677--680.

\bibitem[TCW{\etalchar{*}}95]{tsot95}
\textsc{Tsotsos J.~K., Culhane S.~M., Winky W. Y.~K., Lai Y., Davis N., Nuflo
  F.}:
\newblock Modeling visual attention via selective tuning.
\newblock \emph{Artificial Intelligence 78}, 1-2 (1995), 507--545.

\bibitem[War02]{ward02}
\textsc{Ward M.~O.}:
\newblock A taxonomy of glyph placement strategies for multidimensional data
  visualization.
\newblock \emph{Information Visualization 1}, 3-4 (2002).

\bibitem[War04]{ware04perceptiondesign}
\textsc{Ware C.}:
\newblock \emph{Information Visualization, Second Edition: Perception for
  Design (Interactive Technologies)}.
\newblock Morgan Kaufmann Publishers Inc., 2004.

\bibitem[War08a]{ward08}
\textsc{Ward M.~O.}:
\newblock Multivariate data glyphs: Principles and practice.
\newblock In \emph{Handbook of Data Visualization} (2008), Chen C.-H., Hardle
  W., Unwin A., (Eds.), Springer Handbooks Comp. Statistics. Springer,
  pp.~179--198.

\bibitem[War08b]{ware08visualThinking}
\textsc{Ware C.}:
\newblock \emph{Visual Thinking: for Design}.
\newblock Morgan Kaufmann Publishers Inc., San Francisco, CA, USA, 2008.

\bibitem[WGK10]{ward10interactive}
\textsc{Ward M.~O., Grinstein G., Keim D.}:
\newblock \emph{Interactive Data Visualization: Foundations, Techniques and
  Applications}.
\newblock A K Peters/CRC Press, 2010.

\bibitem[WMM{\etalchar{*}}02]{westin02processing}
\textsc{Westin C.~F., Maier S.~E., Mamata H., Nabavi A., Jolesz F.~A., Kikinis
  R.}:
\newblock {Processing and visualization for diffusion tensor MRI}.
\newblock \emph{Medical Image Analysis 6}, 2 (2002), 93--108.

\bibitem[YPWR03]{yang03}
\textsc{Yang J., Peng W., Ward M.~O., Rundensteiner E.~A.}:
\newblock Interactive hierarchical dimension ordering, spacing and filtering
  for exploration of high dimensional datasets.
\newblock In \emph{IEEE Smyposium on Information Visualization} (2003), IEEE
  Computer Society.

\bibitem[ZSAL08]{zudilova08}
\textsc{Zudilova-Seinstra E., Adriaansen T., Liere R.~v.}:
\newblock \emph{Trends in Interactive Visualization: State-of-the-Art Survey},
  1~ed.
\newblock Springer Publishing Company, Incorporated, 2008.

\end{thebibliography}

\end{document}